\documentclass[12pt]{article}
\usepackage{blindtext}
\usepackage{sbc-template}

\usepackage{graphicx,url}

\usepackage[english]{babel}   
\usepackage[T1]{fontenc} 
\usepackage[utf8]{inputenc} 
\usepackage{multicol}
\usepackage{multirow}
\usepackage{graphicx,url}
\usepackage{amsmath,amssymb,amsfonts}
\usepackage{algorithmic}
\usepackage{textcomp}
\usepackage{url}
\usepackage{enumerate}
\usepackage{multirow}
\usepackage{soul}
\usepackage{verbatim}
\usepackage[table,xcdraw]{xcolor}
\usepackage[shortlabels]{enumitem}
\usepackage{scalefnt} 
\usepackage{comment} 
\usepackage{lscape}
\usepackage{fancyhdr}
\usepackage{sbgames}
\usepackage[super]{nth}
\usepackage[num]{abntex2cite}
\usepackage{hyperref}

\emergencystretch=\maxdimen
\trilha{Indústria} 
\local{Salvador/BA} 
\ano{2025} 
\edicao{XXIV} 

\title{DEFINING THE GAME PRODUCER
A MAPPING OF KEY CHARACTERISTICS AND DIFFERENTIATORS OF THE PROFESSIONAL BEHIND DIGITAL GAME PRODUCTION \\
}
\author{Rafael C. Lopes\inst{1}, Danilo M. Ribeiro\inst{1,2}\ }

\address{CESAR School - CENTRO DE ESTUDOS E SISTEMAS AVANÇADOS DO RECIFE
  \\
  Recife – PE – Brazil
  \nextinstitute
  ZUP INNOVATION\\
  São Paulo – SP – Brazil
  \email{rcl2@cesar.school, dmr@cesar.school}
}

\begin{document} 

\maketitle

\newcommand{\revisao}[1]{\textcolor{red}{ 
  Danilo rev:  #1}}

\thispagestyle{plain}

\begin{abstract}
\textbf{Introduction:} As digital games grow in complexity, the role of the Game Producer becomes increasingly relevant for aligning creative, technical, and business dimensions. 
\textbf{Objective:} This study aimed to identify and map the main characteristics, skills, and competencies that define the Digital Game Producer profile. 
\textbf{Methodology:} A qualitative investigation was conducted with 11 semi-structured interviews, analyzed through Grounded Theory to build categories grounded in professional practice. 
\textbf{Results:} The study produced a structured set of personal characteristics, practical skills, and strategic competencies considered essential for Game Producers. Communication, adaptability, and project management emerged as central elements across the sample. 
\textbf{Conclusion:} The resulting model offers a foundation for professional training, recruitment strategies, and future research on leadership roles in game development.

\end{abstract}


\section{Introduction}

The digital games industry has established itself as one of the most promising sectors of the creative economy, with projections exceeding 200 billion dollars in global revenue by 2024 and an estimated base of over 3 billion players \cite{NEWZOO2}. Beyond entertainment, games are having an increasing impact on areas such as education, healthcare, and marketing, further expanding their social and economic relevance. In this dynamic and constantly evolving landscape, the role of the Game Producer stands out as essential to project success, acting in the coordination of teams and the alignment of technical, creative, and strategic aspects of game development.

Game production differs from conventional software development due to its hybrid nature: it demands not only technical efficiency but also aesthetic sensitivity, narrative understanding, and emotional engagement \cite{Keith}. Furthermore, the development cycle is often iterative and volatile, with frequent changes in scope, mechanics, and product direction throughout the process. As noted by \cite{CHANDLER}, the producer must navigate this unstable environment, reconciling the diverse interests of designers, programmers, artists, testers, and stakeholders—always with a focus on delivering a cohesive, functional, and engaging product. Today, the Game Producer is recognized for their ability to manage schedules, resources, and teams, and is frequently acknowledged for their importance in industry awards such as the D.I.C.E. Awards \cite{TOTTEN}.

Historically, the role of the producer emerged informally, often being performed by programmers or designers themselves during the early stages of the industry. With the professionalization of the sector and the increasing complexity of projects, the position became more specialized, requiring a distinct set of competencies. However, despite the recognized importance of the Game Producer role, no study has been identified that aims to define the specific characteristics, skills, and competencies that constitute an effective Game Producer.

The absence of studies that clearly and systematically define the expected competencies of a successful Game Producer has implications for both the industry and the professional training in this field. Without consolidated references, it becomes difficult to establish objective criteria for the selection, evaluation, and development of these professionals, which may compromise team efficiency and the quality of the final products. Moreover, the lack of academic and practical guidelines hinders the creation of educational curricula and training programs that are aligned with the actual demands of the market. In a sector marked by high competitiveness and rapid technological evolution, this gap may contribute to misalignments between organizational expectations and individual capabilities, ultimately hindering the maturation of the Game Producer’s role as a strategic leader within game development projects.

In light of this context, the following research question was formulated: \textbf{RQ: What are the main characteristics, skills, and competencies that define the role of the Game Producer?} To address this question, the study mapped the most valued attributes in the industry through interviews with active professionals and qualitative analysis of their accounts. By identifying these elements, the research aimed to construct a solid framework, grounded in a rigorous methodological process, capable of guiding the improvement of practices within the digital games sector, as well as contributing to the advancement of academic literature on the subject.

\section{Theoretical Background}
\subsection{Definitions of Characteristics, Skills, and Competencies}




Understanding the concepts of characteristics, skills, and competencies provides the theoretical foundation for identifying what makes a Game Producer effective in complex and multidisciplinary contexts. In this study, we define them as follows:

\textbf{Characteristics} refer to individual traits that influence behaviors, decisions, and interpersonal interactions \cite{Colquitt}. In the context of social skills, they include emotions, beliefs, and ethical choices mobilized in relational situations \cite{DEL_PRETTE}.

\textbf{Skills} are acquired capacities that enable individuals to perform tasks effectively, combining knowledge, practice, and cognitive, physical, and social aptitudes \cite{ARMSTRONG}. They play a central role in both technical and organizational performance.

\textbf{Competencies} integrate knowledge, skills, and attitudes strategically applied to achieve results. According to \cite{DEL_PRETTE}, they represent the articulated mobilization of different forms of knowledge. \cite{FLEURY} defines competencies as the ability to act responsibly, integrating diverse resources in professional contexts.

\subsection{Game Producer}





The Game Producer is identified as a central figure in mediating between technical, creative, and executive domains, being responsible for aligning teams, processes, and deliverables. \cite{Bosch} describes this role as the “ambassador” of the team and the product, while \cite{TÖRNQVIST} compares the producer to a conductor who ensures synchronization across different workstreams.

Among the technical competencies, proficiency in agile methodologies (such as Scrum and Kanban), the use of tools like Jira, and strategic thinking for task prioritization and risk mitigation are particularly emphasized \cite{Bosch, TÖRNQVIST}. Organization and documentation are also highly valued, as they are essential for project traceability.

In terms of interpersonal skills, empathetic leadership, clear communication, active listening, and conflict resolution stand out \cite{Bosch}. These are complemented by adaptability and a genuine interest in games—attributes especially relevant in dynamic and creative environments.

Thus, the Game Producer emerges as a multifaceted professional who integrates technical, organizational, and human dimensions to ensure the delivery of cohesive and innovative games.

\subsection{Related Work}




Studies such as "What Makes a Great Manager of Software Engineers?" \cite{KALLIAMVAKOU} and "What Makes a Great Software Engineer?" \cite{LI} offer valuable contributions by exploring, within technological contexts, the interpersonal, technical, and leadership competencies required for professionals engaged in team mediation and the resolution of complex problems.

Both studies employed semi-structured interviews and analysis grounded in Grounded Theory, enabling the identification of emergent competencies based on participants’ experiences. Among the findings, active listening, empathy, adaptation to team dynamics, clear communication, and a focus on collaborative outcomes were particularly emphasized. For managers, the ability to create a psychologically safe environment and to remove obstacles for the team emerged as key differentiators; for engineers, proactivity, curiosity, and impact orientation stood out.

These studies provided both methodological and analytical foundations for the present research, guiding the data collection and categorization process. By adapting these approaches to the context of digital game production, this study was able to deepen the understanding of the competencies and characteristics that define the profile of the Game Producer.

\section{Research Design}




This study followed a qualitative and exploratory approach, grounded in Straussian Grounded Theory \cite{STRAUSS}, with the aim of constructing analytical categories based on empirical data obtained through interviews with professionals actively working as Game Producers. A total of 11 semi-structured interviews were conducted with participants selected through purposive and convenience sampling, seeking diversity of experiences and practical relevance.

The interviews were conducted synchronously and remotely, using recording and transcription tools. Inclusion criteria were: \textbf{(i)} being over 18 years of age; \textbf{(ii)} having at least three years of formal experience in the role; \textbf{(iii)} participation in released and recognized games; and \textbf{(iv)} self-identification as a Game Producer. Professionals without direct experience in the role or whose activities were limited to project management unrelated to game production were excluded.

The interviewees held positions such as Associate, Senior Producer, Lead Producer, Feature Producer, and Head of Production, reflecting varying levels of responsibility. Most participants were based in Brazil, with additional representation from the United States and the United Kingdom, which enriched the analysis by incorporating cultural diversity.

\subsection{Recruitment and Selection}



Recruitment was conducted primarily through LinkedIn, supplemented by specialized forums (Discord, Slack) and industry events such as the Game Developers Conference. Invitations were sent individually, outlining the objectives of the study and ensuring anonymity and voluntary participation.

The snowball sampling technique was also employed, with initial participants asked to refer other professionals. This approach expanded the reach of the sample and enabled the inclusion of participants from diverse professional contexts.

\subsection{Interviews}



The interview script, presented in Table \ref{tab:exTable1}, aimed to explore the competencies, experiences, and challenges associated with the role of the Game Producer. This approach allows for an in-depth understanding of individual perceptions while maintaining thematic consistency within the investigation \cite{MINAYO}.

Although widely used, the interview method has limitations, such as the bias inherent in subjective responses \cite{CRESWELL}. There is also the risk of interviewer influence, which requires neutrality and skilled listening \cite{CHARMAZ}.

\begin{table}[!htb]
\centering
\caption{Semi-Structured Interview Guide}
\label{tab:exTable1}
\includegraphics[width=.8\textwidth]{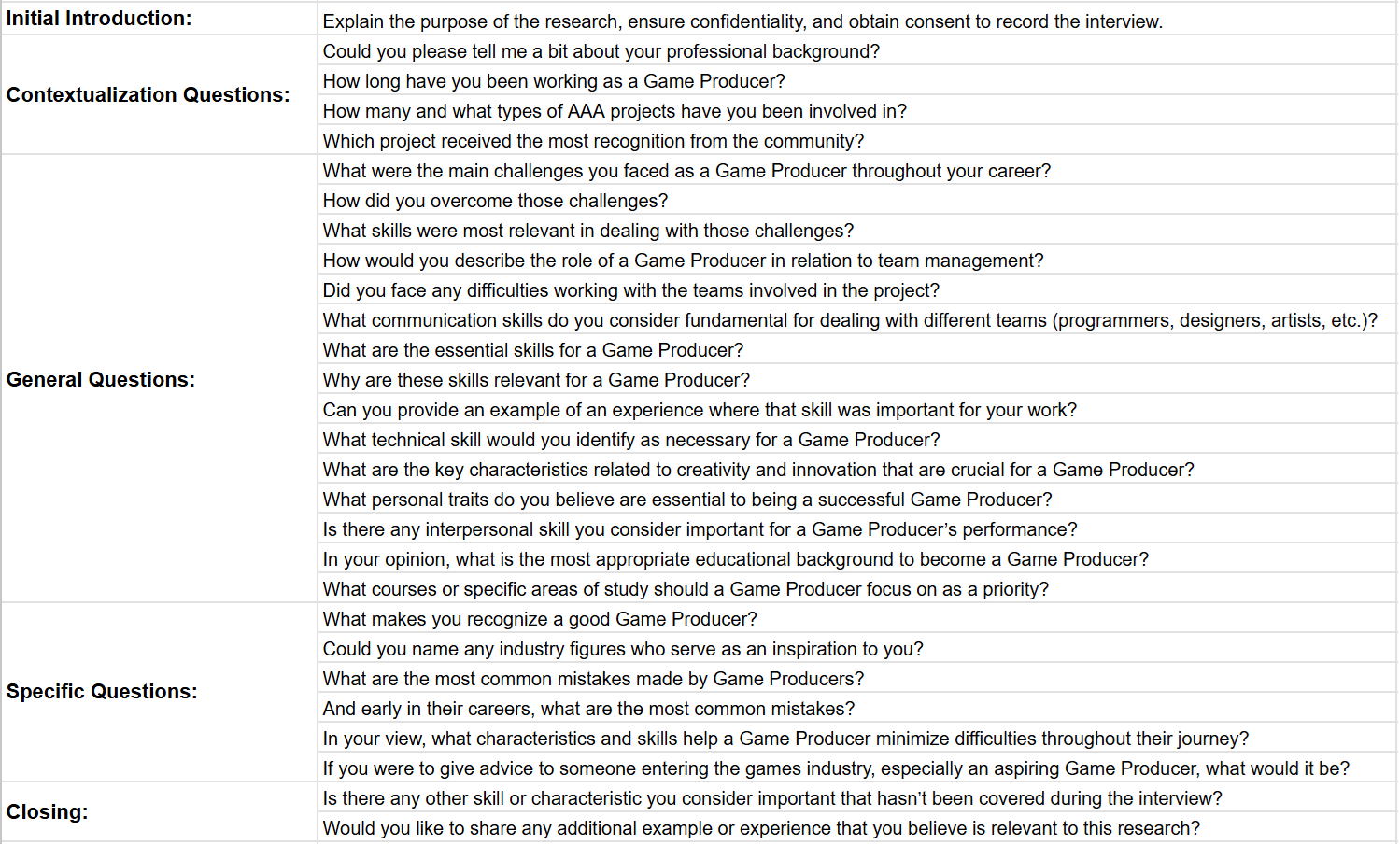}
\end{table}


\section{Results}


Although qualitative analysis prioritizes the interpretation of meanings attributed by participants, the frequency of certain codes may indicate their relevance within the dataset. While count alone does not determine analytical significance, it contributes to identifying patterns and guiding the prioritization of recurring categories \cite{SALDAÑA}.

\subsection{Exemplification  of the Most Representative Codes}





The initial coding process resulted in 195 unique codes, each appearing at least once in the interviews. Among them, “Communication” was the most frequently mentioned, with 94 occurrences in the transcripts. Its definition is presented below, based on a synthesis of participants’ statements and supported by specialized literature.

\textbf{Communication} refers to the ability to convey information clearly, adapt language to the diverse profiles within the team, and foster alignment across multidisciplinary areas. It is essential for ensuring team cohesion and the effective advancement of the project.

\begin{quote}
    \textit{"But, like, I think that if the person knows how to communicate well, they’re going to minimize a lot of problems, you know? ... communicating the risks in the right way, with the right weight, saying the right things to the right people too, ... so I think communication is really key."} (Game Producer 02)
\end{quote}

The following table presents the codes mentioned by more than 60\% of the research participants. This threshold was adopted as an analytical criterion to identify the most recurrent and consensual elements among the interviewees, ensuring significant representativeness within the sample.

To view the dataset used for this analysis, including quotations and codings, please access the \href{https://docs.google.com/spreadsheets/d/1hTB3tJbQAd08H2_gGN5Wu-i12y8AB04U2MYKA04p4MM/edit?gid=573380369#gid=573380369}{link and refer to the "Code Description Quote" tab.}

\begin{table}[!htb]
\centering
\caption{Codes, Definitions and Quotes from Producers}
\label{tab:exTable2}
\includegraphics[width=.9\textwidth]{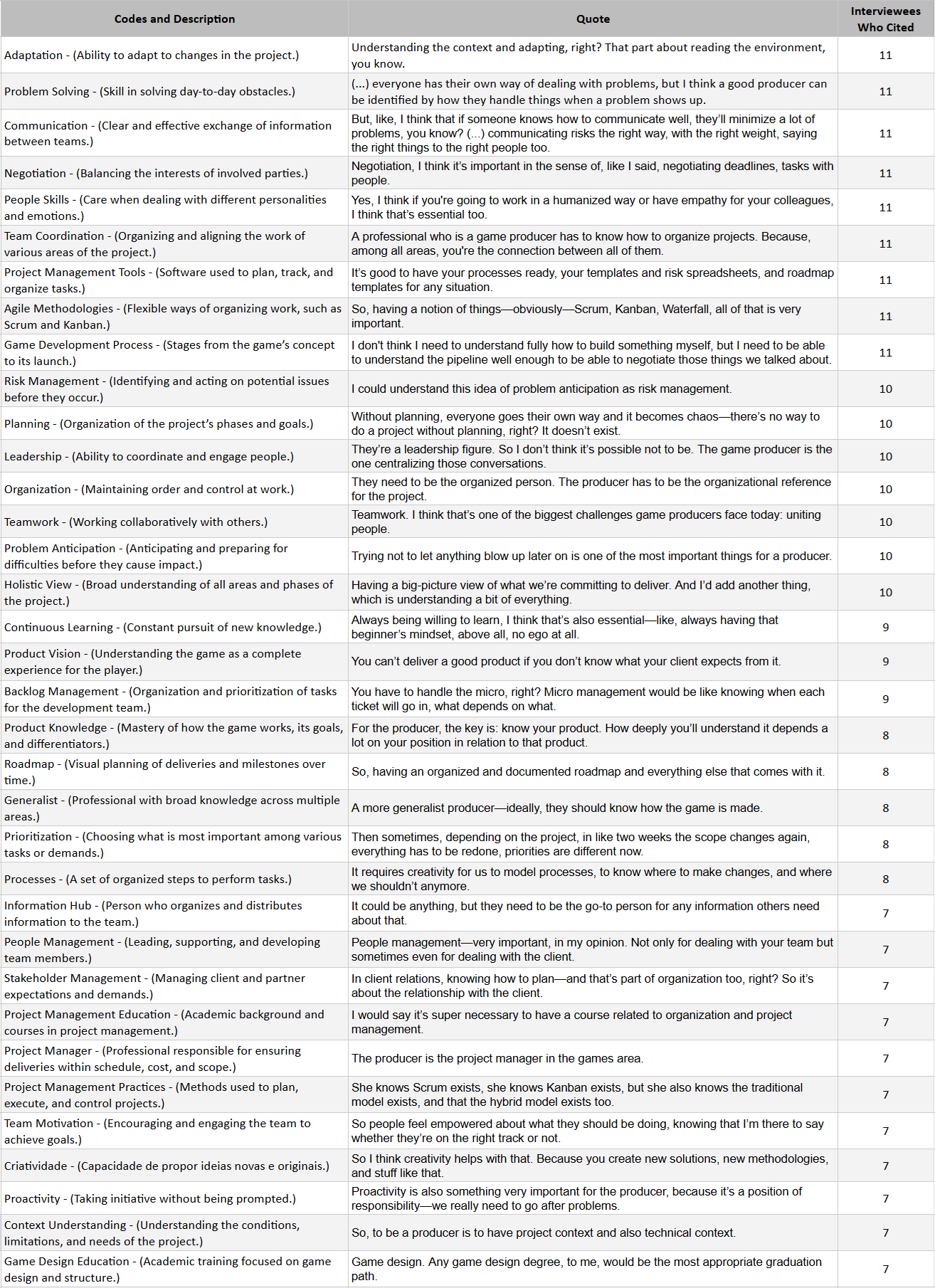}
\end{table}

\subsection{Consolidation into Categories and Subcategories}


The categorization process was carried out by organizing the emerging codes into broader conceptual groups, referred to as categories and subcategories. Each code was associated with a set of recurring meanings identified in the participants' statements, which were then grouped based on thematic and functional similarities. This process enabled the analysis to be structured in a manner consistent with the research objectives, highlighting relevant patterns and facilitating the interpretation of the collected data.


The characteristics of the Game Producer can be understood as relatively stable personal traits that underpin their way of acting, thinking, and interacting in the workplace, especially in contexts of high technical and creative complexity. These characteristics reflect behavioral predispositions that directly influence how the producer perceives themselves and is perceived by others, how they integrate into teams, and how they make decisions amid uncertainty and pressure. With similar representations, the most relevant subcategories of characteristics were "Openness to Knowledge and Development," "Cognitive and Intellectual Skills," and "Professional and Productive Skills."



The skills of the Game Producer refer to the ability to perform concrete actions in the daily routine of digital game development. These skills involve both technical and operational aspects and are generally observable in practice, acquired through professional experience, education, and real-world problem-solving. The subcategory "Communication and Relationship" was the most representative in the coding process. All interviewees emphatically highlighted the importance of communication in the role of the Game Producer. Notable aspects include negotiation, sensitivity in dealing with people, active listening, among others.



The competencies of the Game Producer refer to the ability to mobilize knowledge, skills, and attitudes to effectively address the specific challenges and demands of digital game development. Unlike characteristics, which are related to more stable personal traits, competencies are developed throughout the professional journey and are expressed in the practical application of technical, managerial, and interpersonal knowledge. With equally strong representation, the subcategory of "Project Management" was identified as crucial to the Game Producer’s role. Key aspects include planning ability, risk management, and the knowledge and application of project management practices.


To view all codes associated with the categories, access the \href{https://docs.google.com/spreadsheets/d/1iAx0zR1bDOkOVBN-3f3iqB1u6Z4iBZNPPhjN34EnMts/edit?usp=sharing}{Link, tab "Categories"}

\begin{table}[!htb]
\centering
\caption{Grouping of Codes}
\label{tab:exTable3}
\includegraphics[width=.9\textwidth]{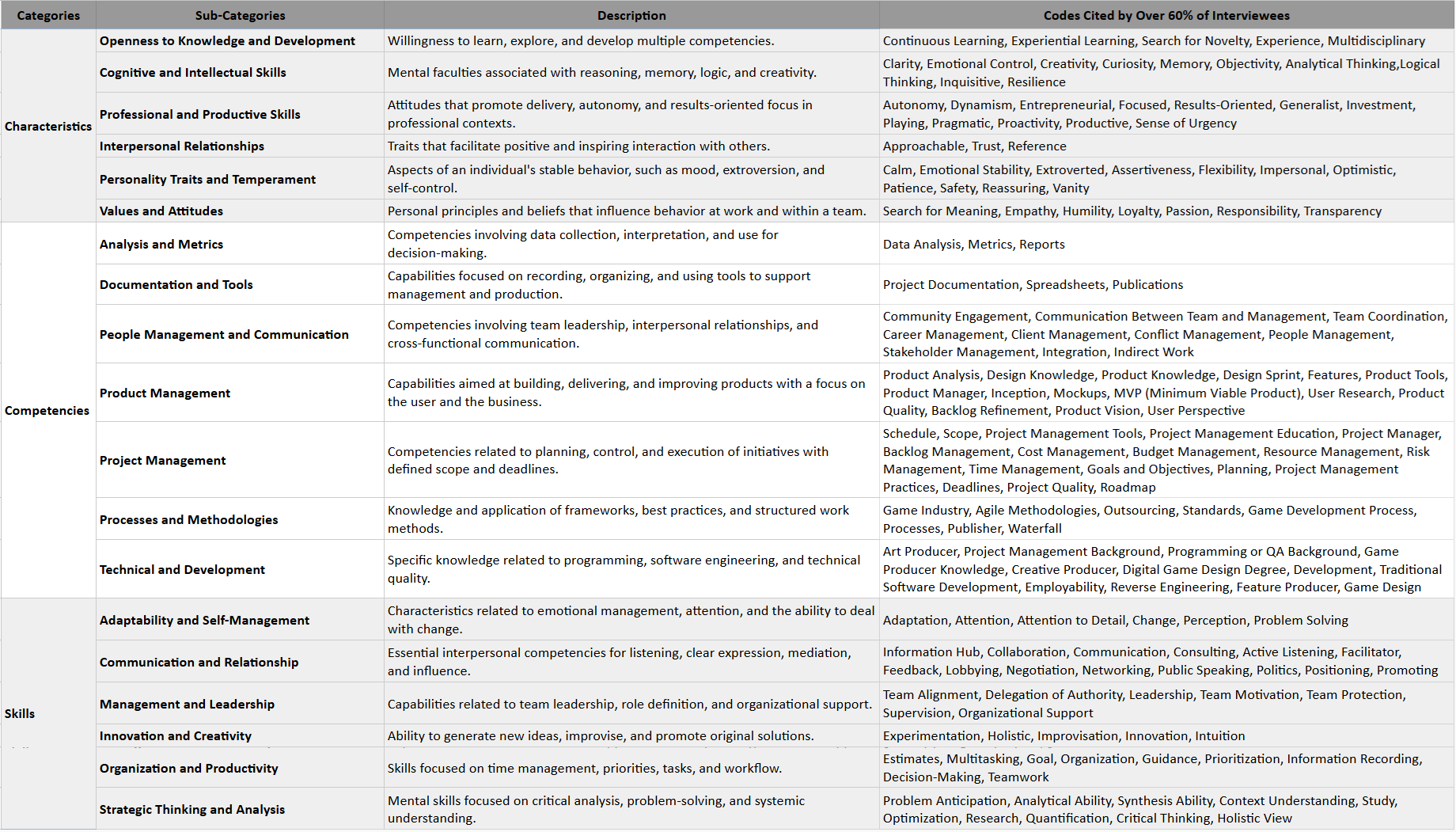}
\end{table}

\section{Threats to Study Validity}





Despite the consistent methodological design and the adoption of the principles of Grounded Theory, this study presents potential threats to validity that warrant consideration. Regarding external validity, the use of convenience and snowball sampling may have limited the diversity of profiles, favoring professionals who are more accessible or visible in networks. As a mitigation measure, diversity in seniority, location, and types of studios was intentionally sought.

Concerning internal validity, there is a risk of researcher bias in the coding and categorization of data, which is inherent to the qualitative approach. To reduce this risk, the study followed the principles of Straussian Grounded Theory and employed digital tools for data organization and traceability.

In terms of transferability, there is the challenge of representing the real practice of the Game Producer without reducing it to idealized notions. To address this, priority was given to preserving participants’ language and contextualizing their statements.

Finally, the risk of premature theoretical saturation must be acknowledged. Although patterns were repeated, the decision was based on interpretative analysis. Future studies may expand the sample and test the model in different contexts.

\section{Discussions}

This study offers valuable contributions to the understanding of the Game Producer’s role, both in practical and academic contexts. From a professional perspective, mapping the expected competencies, skills, and characteristics provides a useful framework for improving recruitment processes, training strategies, and the structuring of multidisciplinary teams. In particular, the emphasis on effective communication, empathetic leadership, and mastery of agile practices suggests that the Game Producer acts as a catalyst for organizational cohesion, operating at the intersection of technical, creative, and business domains.

When compared to the findings of \cite{KALLIAMVAKOU} e \cite{LI}, there is a notable overlap with leadership roles in the technology sector—especially regarding soft skills and team mediation. However, the Game Producer presents a unique profile due to the need to navigate creative volatility, aesthetic sensitivity, and the emotional dynamics of the team—elements less prominent in software engineering contexts. As such, this study broadens the understanding of hybrid roles that do not fit neatly into traditional categories of project management or technical leadership.

Academically, the use of Grounded Theory helps fill a significant gap in the literature, offering a conceptual framework built from empirical industry data. This model may serve as a foundation for future comparative research on other digital production roles, such as Product Managers or Scrum Masters, as well as longitudinal studies on how these competencies evolve throughout a professional career.

The wide range of competencies required—from specific technical skills to the ability to foster team well-being—can lead to role overload or ambiguity, particularly in smaller studios. This complexity underscores the importance of organizational contexts that recognize and support the Game Producer’s role, avoiding its dilution or confusion with other functions.

From a social perspective, by highlighting the value of the Game Producer as an agent of coordination, listening, and innovation, this study contributes to the recognition of collaborative practices within the games industry. Additionally, the findings can inform training policies that are more aligned with real industry demands, promoting greater diversity of profiles and broader access to this expanding field.

At the industry level, the results support the professionalization of the sector by providing a systematic mapping of competencies that foster leadership, effective management of multidisciplinary teams, and the delivery of innovative products.

The study also reinforces the strategic role of the Game Producer as a link between creative, technical, and executive areas, offering recommendations for improvements in recruitment, training, and organizational structure. By integrating qualitative data into the analysis of complex contexts, it contributes to the broader discourse on creative leadership, agile methodologies, and management in digital environments.

Furthermore, the findings provide input for the redesign of curricula in programs related to game production and the creative economy. Socially, the study elevates the profession, highlights its importance in strengthening the creative economy, and may support initiatives aimed at training and deepening the understanding of the profile and needs of this professional. By emphasizing the relevance of digital games as tools for cultural, educational, and social transformation, the research underscores the potential of the Game Producer as an agent of innovation and impact across multiple dimensions.

\section{Final Considerations}

This study aimed to understand the competencies, skills, and characteristics that define the role of the Game Producer in the digital games industry. Through qualitative analysis of interviews with active professionals, it was possible to outline a multifaceted profile that integrates personal attributes, operational capabilities, and strategic competencies. This configuration distinguishes the role from traditional management positions, such as that of project managers.

The results indicate that the Game Producer's performance requires more than mastery of processes and tools. Traits such as resilience, proactivity, and clarity support their professional stance, while practical skills enable leadership, effective communication, and decision-making. Competencies, in turn, synthesize technical knowledge and experience applied to solving complex challenges in game development.

Continuous learning and hands-on experience emerge as central drivers of development. Knowledge in Project Management provides a foundational structure, and effective communication stands out as a key transversal competency for coordinating across teams.

Based on the data, a competency model was developed that can guide recruitment, training, and evaluation practices for Game Producers, while also contributing to the academic literature on game production.

Future directions include: the typification of profiles according to different work contexts; comparisons with related roles such as Product Manager; and investigations into diversity, organizational impact, and adaptation to technological change.

In a constantly evolving industry, understanding the Game Producer also means understanding the dynamics that shape the future of digital games.

\bibliographystyle{sbc}
\bibliography{Referencias/referencias.bib}

\end{document}